\PassOptionsToPackage{unicode}{hyperref}
\PassOptionsToPackage{hyphens}{url}
\documentclass[]{article}
\usepackage{amsmath,amssymb}
\usepackage{lmodern}
\usepackage{iftex}
\usepackage[a4paper, left=2.5cm, right=2.5cm, top=2cm, bottom=2cm] {geometry}
\ifPDFTeX
  \usepackage[T1]{fontenc}
  \usepackage[utf8]{inputenc}
  \usepackage{textcomp} 
\else 
  \usepackage{unicode-math}
  \defaultfontfeatures{Scale=MatchLowercase}
  \defaultfontfeatures[\rmfamily]{Ligatures=TeX,Scale=1}
\fi
\IfFileExists{upquote.sty}{\usepackage{upquote}}{}
\IfFileExists{microtype.sty}{
  \usepackage[]{microtype}
  \UseMicrotypeSet[protrusion]{basicmath} 
}{}
\makeatletter
\@ifundefined{KOMAClassName}{
  \IfFileExists{parskip.sty}{%
    \usepackage{parskip}
  }{
    \setlength{\parindent}{0pt}
    \setlength{\parskip}{6pt plus 2pt minus 1pt}}
}{
  \KOMAoptions{parskip=half}}
\makeatother
\usepackage{xcolor}
\usepackage{longtable,booktabs,array}
\usepackage{calc} 
\usepackage{etoolbox}
\makeatletter
\patchcmd\longtable{\par}{\if@noskipsec\mbox{}\fi\par}{}{}
\makeatother
\IfFileExists{footnotehyper.sty}{\usepackage{footnotehyper}}{\usepackage{footnote}}
\makesavenoteenv{longtable}
\usepackage{graphicx}
\makeatletter
\def\maxwidth{\ifdim\Gin@nat@width>\linewidth\linewidth\else\Gin@nat@width\fi}
\def\maxheight{\ifdim\Gin@nat@height>\textheight\textheight\else\Gin@nat@height\fi}
\makeatother
\setkeys{Gin}{width=\maxwidth,height=\maxheight,keepaspectratio}
\makeatletter
\def\fps@figure{htbp}
\makeatother
\providecommand{\tightlist}{%
  \setlength{\itemsep}{0pt}\setlength{\parskip}{0pt}}
\ifLuaTeX
  \usepackage{selnolig}  
\fi
\IfFileExists{bookmark.sty}{\usepackage{bookmark}}{\usepackage{hyperref}}
\IfFileExists{xurl.sty}{\usepackage{xurl}}{} 
\hypersetup{
  hidelinks,
  pdfcreator={LaTeX via pandoc}}

\author{}
\date{}

\begin{document}

\begin {center}
\section{Maths for Einstein's Universe - \\
Tools for Understanding Modern Reality.}
\end {center}

Anastasia Popkova \textsuperscript{a}, David Blair \textsuperscript{a}
and David Treagust \textsuperscript{b}

Corresponding author: Anastasia Popkova
(\href{mailto:anastasia.popkova@uwa.edu.au}{\nolinkurl{anastasia.popkova@uwa.edu.au}})

\textsuperscript{a} Department of Physics, The University of Western
Australia, Perth WA 6009, Australia

\textsuperscript{b} STEM Education Research Group, School of Education,
Curtin University, Perth WA 6102, Australia

\textbf{Keywords:} mathematics and science education, curriculum
development, Einsteinian physics

 {\subsection {Abstract}

{\small Aversion to mathematics is a recognised and widespread problem.
Following a review of the literature on this subject, this paper
presents an education program which has been developed to test the
hypothesis that transferring attention from traditional school
arithmetic to a broad range of mathematical skills relevant to modern
science at an early age (ages 7-12) will improve students' attitudes to
mathematics, reduce the incidence of maths anxiety and prepare students
for topics normally introduced at more senior levels. The program
entitled \emph{Maths for Einstein's Universe} includes five modules
covering extreme numbers, estimation, probability, vectors and curved
space geometry taught through group activities, games and plays. The
modules complement appropriate early learning of modern physical
concepts from the subatomic world to cosmology. While connected to
science, the program aims to provide meaning and comprehension for
socially relevant topics from national budgets to pandemics and opinion
polls. The program has been trialled in multiple short workshops and
extended learning programs as well as training programs for school
teachers. Analysis of knowledge and attitude tests and questionnaires
from about 170 participants demonstrate strong student enthusiasm and
positive learning outcomes in areas normally considered beyond the
ability of students in this age group. Trial results were used to
identify strategies for enhancing school mathematics based on creation
of stronger links between mathematics and science. We summarise results
of pilot trials. In the paper we present the results of learning powers
of ten and vectors. In total, around 700 participants have trialled
Maths Einstein's Universe with nearly 200 hours of teaching for students
and teachers.}

\textbf{Introduction}

\emph{{Einstein-First project.}} In the early
20\textsuperscript{th} century discoveries in physics created a new
understanding of space, time, matter and radiation, described as
Einsteinian physics. It has given us knowledge of the smallest-scale
phenomena in the universe through to the vast distances of the visible
universe, and with it, the modern technologies on which our lives
depend, and with which we perceive the universe. The Einstein-First
project is an educational response to this revolutionary change in human
understanding. It aims to develop a modernised school science curriculum
so that all students by year 10, their final year of compulsory science
education in Australia, achieve a basic understanding of our current
best understanding of physical reality. An essential component of this
understanding is mathematics.

The distinguished Harvard and Oxford educator Jerome Bruner stated,
``Any subject can be taught effectively in some intellectually honest
form to any child at any stage of development'' (Bruner, 1976). In the
spirit of Bruner, this project is designed to determine an optimum
approach and sequence for meeting our goal. By necessity, the program
must be powerful, fun and mind-expanding. It is based on numerous
peer-reviewed published results.

Many studies have demonstrated that Einsteinian Physics can be
successfully integrated into a curriculum at any level of schooling
{[}Pitts, 2014; Kaur (Part 1-3, 2017; Choudhary, 2018; Choudhary, 2019;
Foppoli, 2019; Choudhary, 2021{]} when taught at the appropriate level
using modern approaches including activity-based learning with toys and
models.

One component that has not been emphasised in the above research is the
development of mathematical concepts that need to be developed in
parallel with the concepts of Einsteinian physics. The need for
modernised mathematics is partly motivated by physics trials where
limitations and contradictions arose because of inadequate mathematical
knowledge, but also by the fact that the same mathematical concepts are
highly relevant to modern life like pandemic prognosis or financial
risk. For this reason, in parallel with the fundamental modernisation of
science education, Einstein-First introduced a goal of re-thinking
mathematics education. The program is entitled \emph{Maths for
Einstein's Universe (MEU).}

Aiming to make the mathematics needed for describing physical reality
more intuitive, more relevant to students' experience, and less
dependent on rote learning, we ask whether Fermi estimation {[}Philip{]}
might be more important than times tables, and whether the conceptual
understanding and representation of scale might be more important than
arithmetic.

\emph{Maths aversion and Maths anxiety.} Math anxiety
(Commodari, 2016) is recognised as a widespread issue for all age
groups. For example, approximately 93\% of adult US Americans indicated
that they experience some level of math anxiety (Luttebberger,
2018).~Maths aversion and anxiety are widely investigated psychological
problems (Stodolsky, 1985), which are not only affecting the
psychological well-being of children (Winarso, 2019) but also affect
students' performance at school (Passolunghi, 2017). The Programme for
International Student Assessment (PISA) studies in 2012 investigated
Maths anxiety amongst 15--16-year-old students in parallel with their
maths performance. Researchers found a strong negative correlation
between students' achievements and reported math anxiety (Luttebberger,
2018).~ Furthermore, this correlation remained stable for several
assessment periods (Pisa 2012; Lee, 2009). The consequences of math
anxiety among students include avoiding math and math-related subjects
in school, as well as aversion to pursuing STEM professions in the
future. (Ashcraft, 2007).

One of the effective strategies to confront maths anxiety was reported
as hands-on and whole-body learning (Thuneberg, 2016). Researchers
argued that human cognitive processes depend on physical perception and
consequently, body movement (Valentini, 2019). A well-known historical
example is fingers-counting. Using the body as an effective tool for
understanding reality makes advanced mathematical concepts less
abstract. Whole-body learning has been recognized as a powerful method
for the comprehension of some mathematical concepts such as fractions
(Isbister, 2018). Einstein-First uses whole-body learning to help
minimise maths anxiety.

Another successful strategy (Maass, 2023) for alleviating Maths anxiety
is connecting maths to real-life situations. This makes maths relevant
to students\textquotesingle{} interest so that it seen as a useful tool,
leading to a build-up of confidence. Examples include estimation of
pocket money, mapping, and maths for sporting competitions.

The approach presented here emphasises maths as a tool for the world,
avoiding abstraction by using toys and activities, as well as connecting
it to physics concepts as well as questions of students' interests which
could be the number of galaxies in the Universe, the size of atoms, or
the total mass of humans.

One of the negative factors which contribute to maths anxiety is putting
students in embarrassing situations when they cannot find the answer and
feel unable to solve a problem. Einstein-First uses toys for teaching.
Children are familiar with toys and every child naturally knows how to
play and what to do. It makes every student confident and emotionally
positive about learning. Once a student has improved maths performance,
the positive outcome provides encouragement for future learning.
Einstein-First has shown that learning with toys and activities,
low-achieving students have similar improvements as higher-achieving
students (Kaur, Part 1, 2017). We hypothesise that the improvement will
also apply to the maths component.

According to (Balazer, 2020). female students are more influenced by
maths anxiety than male students. The success of the Einstein-First
program regarding girls\textquotesingle{} performance (Kaur, 2020)
indicates that the same approach may be beneficial for alleviating
female maths anxiety.

The Einstein-First starts its learning progression in Year 3 (7-8 years
of age), a year before Balazar suggests that maths anxiety begins to be
manifested. We hypothesise that activity-based mathematics learning
analogous to the Einstein-First program (Kaur, Part 2, 2017) may help
ensure that students develop a positive attitude to maths, preventing
maths anxiety before it starts.

\textbf{The structure of the program}

To meet the goal of introducing all students to the core understandings
of modern physics by Year 10 we identified five areas in which
mathematical concepts need to be introduced early in schooling. They are
summarised below, first in brief, then each is expanded to explain the
links and the sequence. Finally, we summarise the module content in
table 1. The five areas of understanding are:

1) The development of logarithmical thinking (Mahajan, S. 2018) to
facilitate understanding of scale.
2) The development of estimation skills to allow students to understand
magnitudes without detailed arithmetic.

3) Development of vector thinking to allow students to develop skills in
symbolic representation for understanding the addition of forces,
momentum, waves and quantum probability.

4) The development of probabilistic thinking (Johnson-Laird, 1994) to
facilitate understanding of the intrinsic probabilistic nature of the
quantum world. 

5) Geometry of curved space to allow students to understand Einsteinian
gravity and comprehend gravitational lensing images and time dilation.
\begin{longtable}[]{@{}
  >{\raggedright\arraybackslash}p{(\columnwidth - 4\tabcolsep) * \real{0.1596}}
  >{\raggedright\arraybackslash}p{(\columnwidth - 4\tabcolsep) * \real{0.2110}}
  >{\raggedright\arraybackslash}p{(\columnwidth - 4\tabcolsep) * \real{0.6294}}@{}}
\toprule()
\begin{minipage}[b]{\linewidth}\raggedright
\textbf{Module name}
\end{minipage} & \begin{minipage}[b]{\linewidth}\raggedright
\textbf{Child-friendly name}
\end{minipage} & \begin{minipage}[b]{\linewidth}\raggedright
\textbf{Module Description}
\end{minipage} \\
\midrule()
\endhead
Module 1. Extreme numbers & Powers of the Universe & This module uses
activities to introduce students to the world of huge and tiny numbers
from the size of atoms to the age of the universe. Powers of two
activities provide a step from linear thinking to logarithmic thinking.
The transition to powers of ten reinforces and extends students'
logarithmic mental number line, allowing the powers of ten notational
tool to be used to conceptualise the extreme scale range of the
universe. Students develop comparison and estimation skills for extreme
numbers, as well as learning how to multiply and divide these numbers,
making extreme numbers easy manageable with simple arithmetic. They are
empowered by new-found ability to estimate the vast scale of the
universe. \\
Module 2. Estimation & Roughly right is better than exactly wrong &
Fermi's statement \emph{Roughly right is better than exactly wrong}
defines the theme of this module that builds on module 1, emphasising
powers of ten notation as tools for obtaining ``roughly right'' answers
to estimation problems. combining it with statistical sampling concepts.
Students learn how to estimate extreme numbers, such as the number of
\emph{stars in the universe} or \emph{the number of solar photons
hitting their palms.} Through experiment they discover the \(\sqrt{N}\)
rule for sampling which is relevant to both opinion polls and photon
imaging in telescopes. \\
Module 3.
Vectors & Maths of arrows & Students learn to connect physical concepts
with symbolic representation in the form of arrows. They discover
experimentally that the addition of arrows is commutative. Through
whole-body force activities and the graphical addition of arrows, they
recognise that both scalars and vectors are needed to describe the
world. Vector understanding is applied to navigation, acceleration,
forces, magnetism, wind and waves, and classical and quantum
interference, all represented graphically. They are introduced to the
phasor wheel, a rolling wheel toy used to connect wave motion to a
rotating vector. \\
Module 4.
Probability & Maths of chance & From Newtonian certainty to Einsteinian
uncertainty: the intrinsic probabilistic nature of our world is revealed
by considering photons reflecting off windows. We see quantum
uncertainty in beam splitters, light interference and radioactivity --
they are all \emph{chance machines}! This is contrasted with uncertainty
for coin tossing and dice where randomness comes from human errors.
Students play games that reveal the concept of statistical distributions
and do experiments to reveal the \(\sqrt{N}\) rule. Applying this to
extreme numbers, students learn how chance connects to \emph{statistical
certainty}. \\
Module 5.

Curved space & Curved Space & This module generalises geometry to the
real non-Euclidean geometry of spacetime. Students study geometry on
curved surfaces to learn that geometry reveals the shape of space. They
study the properties of straight lines in curved space, discover that
\emph{straight is curved if space is curved,} and discover the violation
of Euclid's third postulate and the general breakdown of Euclidean
geometric formulae. A role-play asks \emph{what is the space we live
in?} and reveals how concepts of space and gravity changed over the
years, culminating in the modern understanding that gravity is curved
spacetime. The tiny magnitudes of curved space errors on Earth are
contrasted with their dramatic effects in astronomical images. \\
\bottomrule()
\caption {\small Five modules for Maths for Einstein Universe provide a mathematical tool for better understanding of Einsteinian physics concepts from theory of relativity and cosmology to the quantum world.}
\end{longtable}
The overarching aim of MEU is to teach students that maths is more than
just numbers and create anticipation of future learning. The modules are
described using child-friendly names (Table 1). Each can be delivered at
higher or lower levels of complexity according to the student age group.
To develop a learning sequence for the five core topics, we took into
account a) the current school mathematical curriculum, b) results from
trials in which we assessed students' ability to conceptualize the
relevant mathematical ideas, c) the mathematics requirements needed to
match and enrich the already developed Einstein-First year 3 to 10
science learning progression, and d) the progression of mathematical
learning from module to module.
The five modules form an interrelated set which naturally forms a
sequence while allowing for repetition, extension and reinforcement of
concepts throughout the program. The connections between the five
modules are illustrated schematically in Figure 1.

\begin{figure}[htbp]
  \centering
  \includegraphics [width=6.76806in,height=3.52917in]{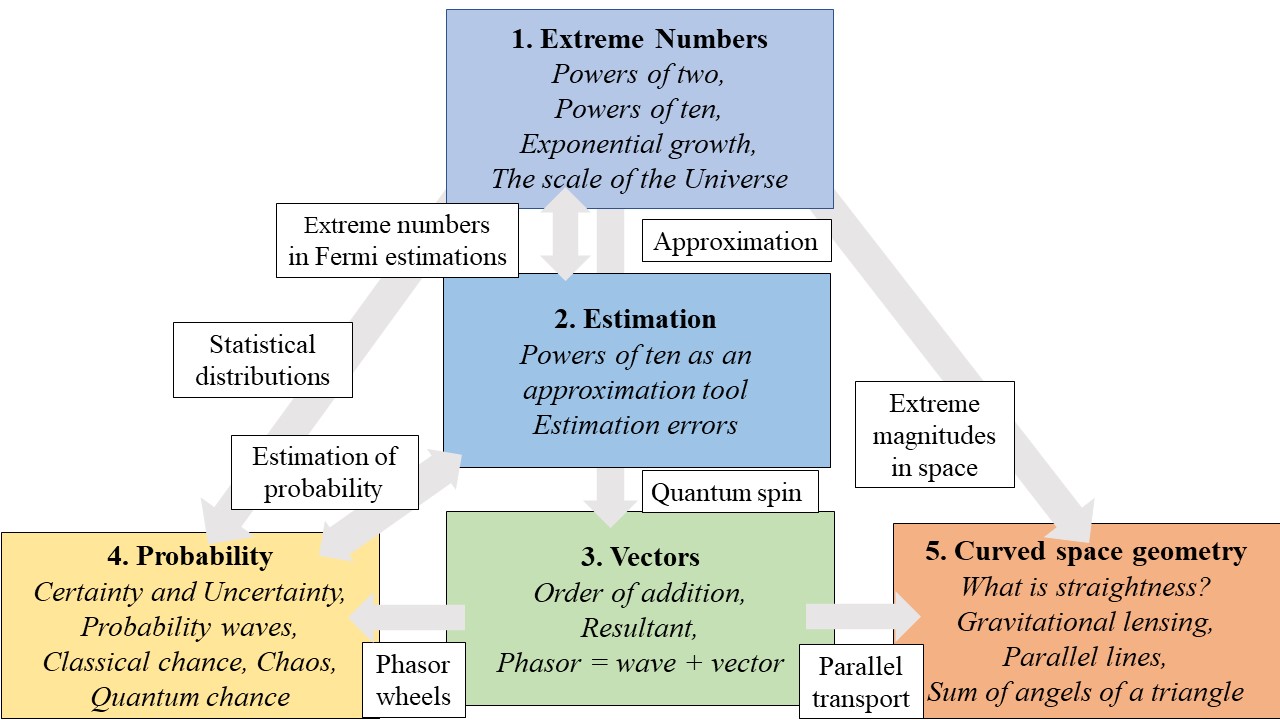}
  \caption{\small {Maths beyond arithmetic; teaching sequence and
interconnections between five modules of Maths of Einstein's Universe.
The grey arrows in-between areas indicate topics, that can be developed
only with pre-knowledge from other modules.}}
  \label{fig:example}
\end{figure}
We start with \emph{Module 1: Extreme Numbers}. All modern areas of
physics involve enormous numbers whether it be cosmological distances or
the number of atoms in a cup of water, as well as the tiny numbers
needed to describe the quantum world. All of these are beyond our direct
perception but are accessible through mathematical reasoning if we first
introduce powers of ten and the logarithmic number line, as described in
Table 1.
The first module leads into \emph{Module 2: Estimations} because powers
of ten notation always involves approximation in the form of rounding.
Approximation is a challenging topic for students for whom mathematics
has focussed on finding single correct answers but powers of ten
exercises in which numbers are approximated to the nearest power of ten
and placed on a logarithmic scale emphasises the connection between
powers of ten and estimation. Once students grasp powers of ten thinking
for understanding the scale of the Universe (\emph{Module 1}), this
skill contributes to Module 2 which is designed to deepen their
understanding of estimations. Students practice power of ten arithmetic
and solve problems with numbers that are inaccessible for exact
calculations or experiments. One example of this sort of problem is to
estimate the number of galaxies in the visible Universe from the Hubble
Deep Field image that shows thousands of galaxies in a tiny patch of sky
(Williams, 2000).

\emph{Modules 1} and \emph{2} focus on numbers, but many quantities are much more
powerfully represented by arrows in the form of vectors, which is the
subject of \emph{Module 3}. Vectors are a higher level representation
where direction and magnitude are equally important. Their use is
familiar in weather forecasting, but many aspects of physical reality,
from forces to magnetism are naturally vector quantities. Also, vectors
can represent the spin of objects, whether they be planets or the
extremely tiny spin of single electrons, which add up to create
macroscopic magnetic forces.

\emph{Module 4}: Probability is focused on the uncertainty of events
(processes) in our universe and connects with \emph{Module 2.} We
identify two types of probability: a) the intrinsic statistical nature
of all quantum processes (quantum probability) such as the chance that a
photon will arrive at a certain location, and b) probability associated
with imprecise knowledge and complex dynamics (classical probability),
such as knowledge of life elsewhere in the universe or the motion of a
rolling dice. As an illustrative and interesting example of mostly
classical probability, we use the Drake equation for estimating the
number of civilisations in the Milky Way galaxy. This connects
probability to the other module topics: estimation and extreme numbers.
We also use videos of single photon interference (Rueckner, 1995) to
allow students to visualise the transition from quantum probability to
\emph{statistical certainty} when the N-value becomes large enough.

\emph{Module 5}: Curved Space extends geometry from the idealised
geometry of Euclid to the actual geometry of curved space, as observed
throughout the universe. The module links both to extreme numbers and to
the understanding of vectors. Extreme numbers connect to the mass range
of observed black holes where spacetime curvature has dramatic and
surprising effects, to the enormous rigidity of space and to the tiny
curvature of space and stretching of time in the weak gravity of the
Earth. The displacement of vector directions when they are
\emph{parallel transported} on curved surfaces allows curvature to be
identified.

\textbf{Maths for Einstein's Universe: activity-based learning.}

The goal of Maths for Einstein's Universe is to give students awareness
of the breadth and scope of mathematics, to open their minds to the
power of mathematics to enhance their understanding of interesting
things, and to discover maths that does not stress the exactness of
traditional arithmetic. Activities with toys and models focus on
tangible physics concepts. For example, as discussed further below, an
activity about photons and beam splitters exposes the statistical nature
of quantum reality while motivating the learning of probability.
Activity-based group learning is free of the stress of obtaining exact
answers and instead creates an enjoyable and memorable experience.

Each lesson begins with an activity that, where possible, involves
multi-sensory learning. By placing the activity first, strong student
engagement is secured, creating the ideal platform for exploring the
mathematical and physical concepts that the activity was designed to
illustrate.

The activities used in MEU can be classified in terms of the dominant
learning instrument involved. We divide them into four categories: a)
the use of toys and models, b) mathematical games, c) role-plays with
songs and poems, and d) the history of discovery. Each of these
approaches provides complementary student experiences involving tactile,
auditory, social and story-based learning. Table 2 gives some examples
in each area.

\begin{longtable}[]{@{}
  >{\raggedright\arraybackslash}p{(\columnwidth - 2\tabcolsep) * \real{0.2762}}
  >{\raggedright\arraybackslash}p{(\columnwidth - 2\tabcolsep) * \real{0.7238}}@{}}
\toprule()
\begin{minipage}[b]{\linewidth}\raggedright
\textbf{Method}
\end{minipage} & \begin{minipage}[b]{\linewidth}\raggedright
\textbf{Examples}
\end{minipage} \\
\midrule()
\endhead
\begin{minipage}[t]{\linewidth}\raggedright
\begin{enumerate}
\def\labelenumi{\arabic{enumi}.}
\item
  Toys and Models
\end{enumerate}
\end{minipage} & Lycra fabric models of orbits in curved spacetime,
spinning tops and gyroscopes for understanding spin vectors, toy atoms
and molecules as tangible representations of extremely small sizes, and
toy cars to define straight line trajectories on curved two-dimensional
spaces, and learning the right-hand rule to find spin vector. \\
\hline
\begin{minipage}[t]{\linewidth}\raggedright
\begin{enumerate}
\def\labelenumi{\arabic{enumi}.}
\setcounter{enumi}{1}
\item
  Mathematical games
\end{enumerate}
\end{minipage} & Exponential dice, a game in which dice sides represent
powers of ten, and scores move the player up or down a logarithmic
number line that extends from -10 (atomic dimensions in meters) to +26
(scale of the visible universe). Rules allow students to develop and
practice their skills and match dimensions to known objects.

A rice grain game on a chess board gives students a tangible
understanding of doublings and powers of two.

The tug of war game explicitly introduces the concept of vector addition
and the idea of the resultant vector. \\
\hline
\begin{minipage}[t]{\linewidth}\raggedright
\begin{enumerate}
\def\labelenumi{\arabic{enumi}.}
\setcounter{enumi}{2}
\item
  Role plays and songs
\end{enumerate}
\end{minipage} & Discovery of Zero is a role-play about a birthday party
set 4000 years ago when zero was invented as a placeholder to prevent
errors in arithmetic (Blair, 2023).

Ten Times Alice is a role-play based on Alice in Wonderland in which
bites of magical food cause Alice to change size ten times. Songs and
imagination reinforce the rules of division and multiplication and allow
students to practice logarithmic thinking (Blair, 2023). \\

\begin{minipage}[t]{\linewidth}\raggedright
\begin{enumerate}
\def\labelenumi{\arabic{enumi}.}
\setcounter{enumi}{3}
\item
  Historical content
\end{enumerate}
\end{minipage} & The history of zero as an operator, which enables
logarithmic thinking and allows us to imagine the entire universe.
The millennia-long quest to determine the value of $\pi~$to greater and
greater precision includes feats of memory and supercomputing.

The history of the understanding of space from Euclid to the precise
observation of curved space at Wallal (Blair, 2022) including Gauss's
mountain top triangle experiment where he explored the understanding of
geometry, the testing of Euclid's geometry and the concept that light
trajectories define straight lines. \\
\bottomrule()
\caption {\small Four main components are employed to construct activity-based
lessons for Maths for Einstein's Universe in a non-abstract and
age-appropriate way by using toys, games, drama, and history.}
\end{longtable}
To further illustrate how activities designed with these learning
instruments can elucidate the mathematical concepts in an
age-appropriate and engaging way, we give three examples. First an
exponential growth activity relating to \emph{Modules 1} and \emph{2}. Second an
activity about vectors in an abstract context (\emph{Module 3}), and finally an
activity on probability for a quantum world (\emph{Module 4}).

1.\emph{The mathematician tricks the emperor}: The
rice-on-the-chessboard activity is a realisation of the well-known story
of the mathematician who tricks the emperor (Demi,1997). Following the
mathematician's demand, students place one grain on the first square and
double the number for each square. Students typically manage to fill
about 10 squares (2\textsuperscript{10} =1024 grains). Having physically
realised the first ten squares, the activity progresses using estimation
and imagination. Students have previously learnt through other activities,
and repeated here, that 2\textsuperscript{10} \textasciitilde{}
10\textsuperscript{3}. They apply this approximation to the chess board,
recognising that 2\textsuperscript{20} \textasciitilde{}
10\textsuperscript{6}, 2\textsuperscript{30}
\textasciitilde10\textsuperscript{9} etc, ultimately determining the
number of rice grains for square number 64 to be \textasciitilde16 x
10\textsuperscript{18} grains.

This physical activity connects powers of two to powers of ten, and
utilises both estimation and extreme numbers, all connected to the
concept of exponential growth. It provides a framework for developing
abstract representation and logarithmic understanding of other domains
such as time and distance. If one rice grain represents one second, the
age of the universe is represented by the square number 60
\textasciitilde10\textsuperscript{18} seconds. If one grain represents 1
meter, then square number 40 takes you to the edge of the Solar system
and square number 54 represents the distance to the nearest star. The
combination of activity, imagination, estimation, and simple arithmetic
has transported the students from pure numbers to astronomical scales of
time and space.

2. \emph{Phasor wheel vectors}. In Module 3, vectors are first
introduced for tangible representations such as forces as described in
Table 2, item 3 (Tug of war game). We also introduce two abstract
representations. First, we used phasor vectors as a tool for
understanding the interference of light. Students observe the striking light
interference patterns when a laser pointer is reflected off a soap film.
The patterns of light and dark are taught as a vector addition process.
A toy called a phasor wheel is introduced, which uses a rolling wheel
with a phasor vector and a cam bar attached. (Choudhary , 2021) It shows
the connection between a rotation, wave motion (the displacement of the
cam) and the phasor (a radial arrow on the wheel). Bright spots arise
when the phasor arrows from two paths are aligned.

Secondly, we used vectors to describe spin. We use the term spin, rather
than angular momentum because we wish to introduce the concept of
quantum spin using the analogy with classical spin. To introduce the
spin vector, we simply introduce the right-hand rule and use simple toy
tops to study the physical concept of spin, including the addition and
cancellation of spin and the precession of spin vectors when forces are
applied.

3. \emph{Quantum probability and partial reflections}: Students
investigate the reflection of a laser pointer from a window or else they
photograph coloured objects seen directly, in a mirror and in a partial
reflection from a window. Pre-knowledge is introduced, that light comes
as photons and their energy depends on their colour. The key observation
made by students is that partial reflections do not change colour --
they only change their intensity. This means that all the photons are
reflected ``whole''. How does each photon "know" where to go?
It\textquotesingle s a matter of quantum chance. For normal window
glass, the probability of a single photon being reflected very roughly
equates to the probability of rolling a double six with a pair of dice.
By rolling dice and tabulating double-sixes versus all other scores,
students obtain an experiential connection between photon reflection,
the statistical nature of quantum physics, and the concept that image
brightness represents the number of photons.
Further activities are presented in (Popkova, 2020), (Blair, 2022) and
the accompanying paper (Popkova, 2023)."

\textbf{Trials of Maths for Einstein's Universe}

The MEU program was developed iteratively following the Model of
Educational Reconstruction in which trial results were used to identify
the most effective learning instruments at different ages. For initial
trials we offered approximately 40 diverse 45 minute out-of-school
workshops covering components from all five modules. They attracted the
participation of over 700 primary and secondary school students.
Following the initial trials, formal lessons were carefully designed and
trialled in schools as both long-term and short-term programs.These trials are
presented in Table 3. 

\begin{longtable}[]{@{}
  >{\raggedright\arraybackslash}p{(\columnwidth - 10\tabcolsep) * \real{0.1110}}
  >{\raggedright\arraybackslash}p{(\columnwidth - 10\tabcolsep) * \real{0.1273}}
  >{\raggedright\arraybackslash}p{(\columnwidth - 10\tabcolsep) * \real{0.1909}}
  >{\raggedright\arraybackslash}p{(\columnwidth - 10\tabcolsep) * \real{0.2549}}
  >{\raggedright\arraybackslash}p{(\columnwidth - 10\tabcolsep) * \real{0.1750}}
  >{\raggedright\arraybackslash}p{(\columnwidth - 10\tabcolsep) * \real{0.1409}}@{}}
\toprule()
\begin{minipage}[b]{\linewidth}\raggedright
\end{minipage} & \begin{minipage}[b]{\linewidth}\raggedright
\textbf{Modules}
\end{minipage} & \begin{minipage}[b]{\linewidth}\raggedright
\textbf{Dates}
\end{minipage} & \begin{minipage}[b]{\linewidth}\raggedright
\textbf{Number of sessions x Duration}
\end{minipage} & \begin{minipage}[b]{\linewidth}\raggedright
\textbf{Number of participants}
\end{minipage} & \begin{minipage}[b]{\linewidth}\raggedright
\textbf{Year level}
\end{minipage} \\
\midrule()
\endhead
\begin{minipage}[t]{\linewidth}\raggedright
\begin{enumerate}
\def\labelenumi{\alph{enumi})}
\tightlist
\item
\end{enumerate}

\end{minipage} 

& 1 and 5 & May 2021.

Nov. 2022

4 weeks & 3 x 1 hr

1 x hr & 28 & 5-9 \\
\hline
\\
\begin{minipage}[t]{\linewidth}\raggedright
\begin{enumerate}
\def\labelenumi{\alph{enumi})}
\setcounter{enumi}{1}
\tightlist
\item
\end{enumerate}
\end{minipage} & 3 & Jul. 2021

Two days & 2 x 2 hr & 27 & 4 \\
\hline
\\
\begin{minipage}[t]{\linewidth}\raggedright
\begin{enumerate}
\def\labelenumi{\alph{enumi})}
\setcounter{enumi}{2}
\tightlist
\item
\end{enumerate}
\end{minipage} & 1 and 3 & Jul. 2021

Two days & 3 x 2 hr & 9 & 5-6 \\
\hline
\\
\begin{minipage}[t]{\linewidth}\raggedright
\begin{enumerate}
\def\labelenumi{\alph{enumi})}
\setcounter{enumi}{3}
\tightlist
\item
\end{enumerate}
\end{minipage} & 1 & Sept 2021

4 weeks & 4 x 1.5 hr, & 27 & 5-6 \\
\hline
\\
\begin{minipage}[t]{\linewidth}\raggedright
\begin{enumerate}
\def\labelenumi{\alph{enumi})}
\setcounter{enumi}{4}
\item
\end{enumerate}
\end{minipage} & 1- 5 & Feb. 2022

-May 2023 & 53 x 2hr & 30 & 2 - 9 \\
\hline

\begin{minipage}[t]{\linewidth}\raggedright
\begin{enumerate}
\def\labelenumi{\alph{enumi})}
\setcounter{enumi}{5}
\tightlist
\item
\end{enumerate}
\end{minipage} & 1, 2, 3, 5 & Jul. 2022.

Sept 2022

Two days & 2 x 1hr

1 x 1hr & 26 & 5-9 \\
\hline
\\
\begin{minipage}[t]{\linewidth}\raggedright
\begin{enumerate}
\def\labelenumi{\alph{enumi})}
\setcounter{enumi}{6}
\tightlist
\item
\end{enumerate}
\end{minipage} & 1 & Aug. 2022

4 weeks & 4 x 1.5 hr & 30 & 5-6 \\
\hline
\\
\begin{minipage}[t]{\linewidth}\raggedright
\begin{enumerate}
\def\labelenumi{\alph{enumi})}
\setcounter{enumi}{7}
\tightlist
\item
\end{enumerate}
\end{minipage} & 3 & May 2023
4 weeks & 16 x 1 hr & 100 & 5-6 \\
\hline
\\
\textbf{Totals} & \textbf{1-5} & \textbf{2021-2023} & \textbf{151 hr} &
\textbf{277} & \\

\bottomrule()
\caption{\small Trials of Maths for Einstein's Universe in 2021-2023: a)
after-school workshops in the Einstein-First facility b), d), g), h)
government schools, c) PEAC program e) long-term program for a stable
multi-age class, taught in three different age groups f) remote
aboriginal school.}
\end{longtable}
Feedback from participants was collected and analysed for the purpose of developing and testing each module. We
collected data from 277 students and 63 teachers. 
We also collected data on teacher's responses to our teacher training
courses that consisted of micro-credential courses (micro-credentials,
2023) and training workshops. Table 4 presents a summary of the teaching
training and micro-credential courses for teachers.\\

\textbf{Observations and Qualitative Outcomes}

The trials described in Table 3 included students across an age range
from 7 to 14. Because the program covered material normally outside the
school curriculum, we had no benchmarks or prior data to calibrate our
expectations. For this reason, observations and qualitative measures
were essential for testing and refining lessons, evaluation of the
optimum age for introducing learning content, and evaluation of student
responses.

\begin{longtable}[htbp]{@{}
  >{\raggedright\arraybackslash}p{(\columnwidth - 10\tabcolsep) * \real{0.0409}}
  >{\raggedright\arraybackslash}p{(\columnwidth - 10\tabcolsep) * \real{0.3831}}
  >{\raggedright\arraybackslash}p{(\columnwidth - 10\tabcolsep) * \real{0.1502}}
  >{\raggedright\arraybackslash}p{(\columnwidth - 10\tabcolsep) * \real{0.1187}}
  >{\raggedright\arraybackslash}p{(\columnwidth - 10\tabcolsep) * \real{0.1558}}
  >{\raggedright\arraybackslash}p{(\columnwidth - 10\tabcolsep) * \real{0.1513}}@{}}
\toprule()
\begin{minipage}[b]{\linewidth}\raggedright
\end{minipage} & \begin{minipage}[b]{\linewidth}\raggedright
\textbf{Name of course/workshop}
\end{minipage} & \begin{minipage}[b]{\linewidth}\raggedright
\textbf{Dates}
\end{minipage} & \begin{minipage}[b]{\linewidth}\raggedright
  \textbf{Duration}
\end{minipage} & \begin{minipage}[b]{\linewidth}\raggedright
  \textbf{Audience}
\end{minipage} & \begin{minipage}[b]{\linewidth}\raggedright
\textbf{Number of participants}
\end{minipage} \\
\midrule()
\endhead
\begin{minipage}[t]{\linewidth}\raggedright
\begin{enumerate}
\def\labelenumi{\alph{enumi})}
\tightlist
\item
\end{enumerate}
\end{minipage} & Maths of arrows & Oct. 2021 & 1 hr & Pre-service
teachers & 20 \\
\begin{minipage}[t]{\linewidth}\raggedright
\begin{enumerate}
\def\labelenumi{\alph{enumi})}
\setcounter{enumi}{1}
\tightlist
\item
\end{enumerate}
\end{minipage} & Maths for Einstein's Universe

(micro-credential courses) & Jan. 2022-

Feb. 2022 & 18 hr & Primary and secondary teachers & 9 \\
\begin{minipage}[t]{\linewidth}\raggedright
\begin{enumerate}
\def\labelenumi{\alph{enumi})}
\setcounter{enumi}{2}
\tightlist
\item
\end{enumerate}
\end{minipage} & Maths for Einstein's Universe

(micro-credential courses) & Aug. 2022-

Oct. 2022 & 18 hr & Primary and secondary teachers & 8 \\
\begin{minipage}[t]{\linewidth}\raggedright
\begin{enumerate}
\def\labelenumi{\alph{enumi})}
\setcounter{enumi}{3}
\tightlist
\item
\end{enumerate}
\end{minipage} & How small is an atom?

Numbers in the Universe on a Chessboard & Oct. 2022 & 1.5 hr & Primary
teachers & 12 \\
\begin{minipage}[t]{\linewidth}\raggedright
\begin{enumerate}
\def\labelenumi{\alph{enumi})}
\setcounter{enumi}{4}
\tightlist
\item
\end{enumerate}
\end{minipage} & Maths for Einstein's Universe

(micro-credential courses) & May. 2023-

Jul. 2023 & 9 hr & Primary and secondary teachers & 15 \\
\multicolumn{2}{@{}>{\raggedright\arraybackslash}p{(\columnwidth - 10\tabcolsep) * \real{0.4240} + 2\tabcolsep}}{%
\textbf{Totals}} & \textbf{2021-2023} & \textbf{46.5 hr} & &
\textbf{63} \\
\bottomrule()
\caption{\small MEU teacher training courses: a) and d) workshops; b), c) and
e) part of a 72-hour micro-credential course.}

\end{longtable}

The qualitative observations are drawn from post-lesson notes for each
workshop as well as some video and photographic data. The MEU activities
were differentiated to suit the various age groups and ability levels,
while in some cases mixed ages allowed direct comparison of age-related
engagement.

The observations have allowed us to obtain the following general
description of student performance and age-related aptitude, providing
data used to define the next iteration of the MEU program. We
categorized our observations using toys, mathematical games, role-plays,
concept engagement duration, language development, age-dependent
learning styles, and learning retention across three main groups of
children: a) Year 3-4, b) Year 5-6, and c) Year 7-9:

\begin{enumerate}
\def\labelenumi{\arabic{enumi}.}
\item
  \emph{Engagement with toys:} All students regardless of age were
  observed to relate readily to activities with toys, such as toy atoms
  and molecules, toy photons and toy spacetime, as well as related
  games. The toys clearly provided visual and tactile meaning for the
  relevant mathematical and physical concepts, such as scale, speed,
  geometry and statistical properties. Students accepted the toy models,
  without appearing to need to question their adequacy as accurate
  representations.
\item
  \emph{Mathematical games.} From youngest to oldest, independent on
  gender, all students enthusiastically participated in mathematical
  games. It was observed that students in years 3-4 showed a higher
  level of engagement and preference for games that involved physical
  and tactile experiences, compared to older year groups.
\item
  \emph{Concept engagement duration}. The introduction of new concepts
  and language requires time for assimilation. The activities with toys
  extended the duration that students across all the age groups engage
  with key concepts, expediting the formation of their cognitive
  concepts on a deeper level.
In relation to this we asked what the optimum duration of activities is,
and how does the optimum duration depend on age. We observed that both
Year 5-6 students and Year 3-4 students showed an equal amount of time
being involved in playing. However, Year 5-6 students were more engaged
with advanced toys such as phasor wheels, whereas year 3-4 students
tended to spend more time with simpler toys like toy atoms. Students in
Year 7-9, while engaging with toys were quickly ready to move on to an
analytical and formal explanation, balancing playing with mathematical
reasoning.

We were surprised to observe that students in Year 3-6 not only achieved
similar scores but sometimes also demonstrated a deeper understanding of
certain concepts, thanks to their prolonged engagement with educational
toys. Older students who spent less time engaging with toys sometimes
missed important detail. For instance, in working with vector addition,
older students who spent less time engaged with toy magnetic arrows were
more likely to confuse the direction of a resultant vector.

It is important to note that an excessive duration of unstructured play
can lose the connection between the toy and the concept. The best
learning outcomes occurred when the connection between the toy and the
concept was repeatedly reinforced by the teacher, and by sequential
activity tasks.

A fraction of students, especially those from lower socio-economic
backgrounds, found it difficult to establish a strong link between
activities and concepts. In a low SES school, one third of students
appeared to experience this difficulty, while in extension classes, only
1-2 students in a typical class (a few percent) faced this challenge.

Despite these learning challenges, the universally high engagement with
the activities, and positive student responses (see below) leads us to
believe that the pleasurable and memorable tactile associations are
likely to contribute positively to future learning.
\end{enumerate}

\begin{enumerate}
\def\labelenumi{\arabic{enumi}.}
\setcounter{enumi}{3}
\item
  \emph{Role Plays}: Role-playing combines historical knowledge with
  imagination and creativity. It was particularly effective for Year 3-5
  school students, who were highly engaged. A common response at the end
  of a role-play session was `can we do it again?'. Songs in the role
  plays were clearly remembered, and very useful for cementing concepts
  such the role of zero, as well as difficult new words such as photon.
One weakness of role-play learning is that drama can take precedence
over conceptual understanding. We found that it was very important for
the teacher to draw-out meaning through classroom discussion and
examination of the concepts.
\end{enumerate}

\begin{enumerate}
\def\labelenumi{\arabic{enumi}.}
\setcounter{enumi}{4}
\item
  \emph{Learning styles.} Students in years 3-4 showed impressive
  ability to connect concepts to everyday experiences and imagination.
  For example, students in this group were easily able to recognize the
  trajectory of a cam-rod on a phasor wheel (Choudhary, 2021) as an
  approximate wave trajectory: \emph{`It is like an ocean wave'}, stated
  an 8-year-old during program e) (table 3).
Year 5-6 students demonstrated a remarkable ability to embrace
scientific notation by only establishing connections between the number
of zeros in a billion and index notation. Formal mathematical
conceptualization of exponentials is clearly not needed to create the
basic skills for logarithmic thinking. Year 7-9 students exhibited a
keen interest in applying powers of ten notation for tackling large
number estimation problems.
\end{enumerate}

\begin{enumerate}
\def\labelenumi{\arabic{enumi}.}
\setcounter{enumi}{5}
\item
  \emph{Learning retention.} In many trials younger (Year 3-4) and older
  children (Year 7-9) demonstrated similar learning performance.
  However, by testing after a few weeks, we found that younger students
  forget more quickly. This reduced level of retention is not surprising
  (Piaget, 1983). In a spiral curriculum where concepts are revisited in
  later years, their early comprehension is still likely to assist in
  developing deeper understanding in later years, but in general this
  observation implies that younger students need ongoing reinforcement.
\end{enumerate}

\textbf{Results and Evaluations}

We shall now present general results obtained from our trials, which
provide a comprehensive expression of the program\textquotesingle s
effectiveness. In addition to formal questionaries, parents and students
were asked simple evaluation questions. Recorded answers were
transcribed and approved for correctness. Teachers provided in-depth
written responses to evaluation questions.

First, we will present some examples of student's attitude results from
across the MEU program, followed by knowledge test results from
individual programs. Then we will go on to present evaluation results
from across the MEU program.

\emph{Attitude questions.} One of our primary objectives of MEU
is to enhance children\textquotesingle s attitudes towards mathematics
and science. In program f, as outlined in Table 3, students engaged in
a chessboard activity as part of the program \emph{The mathematician
tricks the emperor} (see above), a paper cutting activity (Popkova,
2020), maths of arrows linked to hair interference experiments and
eclipse activities (Blair, 2022). In the post-test, two statements were
designed to assess students\textquotesingle{} perceptions of learning
through activities, using a Likert scale. More than 90 percent of
children taking part in Maths for Einstein's Universe agreed or strongly
agreed with the statement, "I like learning things with activities'' and
``Science with activities is fun'' while 73 percent agreed or strongly
agreed with the statement, "I like doing activities to learn about
numbers."

\emph{Knowledge questions.}

\begin{enumerate}
\def\labelenumi{\alph{enumi})}
\item
  The results of pilot trials of \emph{Powers of the Universe} module
  from pre-test and post-test comparisons indicate that nearly 100\% of
  students were able to use powers of two for describing numbers in
  processes such as bacteria splitting or numbers of ancestors. In the
  post-test, 78\% of students preferred to use powers of ten notation
  over multiple zeros for big and small numbers. According to post-test
  results, 98\% of students expanded their understanding of the scale of
  the Universe by demonstrating knowledge of cosmological objects and
  subatomic particles in response to questions about the largest and
  smallest objects or magnitudes they were familiar with.
\item
  The results of the \emph{Maths of Arrows} module from the post-test
  demonstrated that nearly 100\% of students were able to understand the
  idea of vector and vectors addition as well as use it for solving
  problems such as the addition of forces when many people push a car.
\item
  Understanding phasors and quantum spin vectors: In post-testing nearly
  90\% of students were able to correctly use the right-hand rule for
  defining spin vectors of spinning objects. More than 83\% of students
  could add phasors shown by a phasor wheel and proximately 90 \% could
  correctly correlate the resultant with light intensity in an
  interference pattern.
\end{enumerate}

Detailed outcomes of the program can be found in the papers (Popkova,
2020), (Blair, 2022) and forthcoming papers.

\emph{Feedback responses from interviews with teachers, parents,
students}

To assess MEU programs, we collected written feedback from each teacher
(T) who delivered the program or assisted in the teaching, including a
STEM coordinator and a teaching principal. We also collected verbal
feedback (recorded and transcribed) from parents (P), and students (S).
We observed that all responses shared four common parameters given in
table 5. Furthermore, we conducted an analysis of
teachers\textquotesingle{} confidence. We give some typical responses in
Table 5, followed by a more detailed analysis.

\begin{longtable}[]{@{}
  >{\raggedright\arraybackslash}p{(\columnwidth - 2\tabcolsep) * \real{0.2825}}
  >{\raggedright\arraybackslash}p{(\columnwidth - 2\tabcolsep) * \real{0.7175}}@{}}
\toprule()
\begin{minipage}[b]{\linewidth}\raggedright
\textbf{Feedback parameters}
\end{minipage} & \begin{minipage}[b]{\linewidth}\raggedright
\textbf{Examples of feedback (exact quotations)}
\end{minipage} \\
\midrule()
\endhead
\begin{minipage}[t]{\linewidth}\raggedright
\begin{enumerate}
\def\labelenumi{\arabic{enumi})}
\item
  Program structure and way of learning
\end{enumerate}
\end{minipage} & \textbf{T:} \emph{``A good blend between concepts,
activities, games and assessment.''}

\textbf{P:} \emph{``I am truly impressed by the wide range of
activities.''} \\
\hline
\\
\begin{minipage}[t]{\linewidth}\raggedright
\begin{enumerate}
\def\labelenumi{\arabic{enumi})}
\setcounter{enumi}{1}
\item
  Enthusiasm and engagement
\end{enumerate}
\end{minipage} & \textbf{T:} \emph{``Students\textquotesingle{}
attendance remained high even though it was an after-school program.''}

\textbf{P:} \emph{``At home, he always talks about the activities.''}

\textbf{S:} \emph{``It was so cool for all of us to push the car
together!''} \\
\hline
\\
\begin{minipage}[t]{\linewidth}\raggedright
\begin{enumerate}
\def\labelenumi{\arabic{enumi})}
\setcounter{enumi}{2}
\item
  Effectiveness
\end{enumerate}
\end{minipage} & \textbf{T:} \emph{``The review questions asked at the
beginning of the sessions were a further indicator that students were
attending and understanding what had been previously covered.''}

\textbf{P:} \emph{My son had no problems with advanced scientific
problems in school extension classes\ldots\ldots.} \\
\hline
\\
\begin{minipage}[t]{\linewidth}\raggedright
\begin{enumerate}
\def\labelenumi{\arabic{enumi})}
\setcounter{enumi}{3}
\item
  Teachers' confidence
\end{enumerate}
\end{minipage} & \textbf{T:} \emph{``I did need to ask her}
{[}Einstein-First team member{]} \emph{a few times whether what I was
teaching was correct.''} \\
\hline
\\
\begin{minipage}[t]{\linewidth}\raggedright
\begin{enumerate}
\def\labelenumi{\arabic{enumi})}
\setcounter{enumi}{4}
\item
  Future collaboration
\end{enumerate}
\end{minipage} & \textbf{T:} \emph{``I look forward to teaching it
again.''}

\textbf{P:} \emph{``She wants to learn science and maths in this way.''}

\textbf{S:} \emph{``I want to play with toys to learn maths and
science.''} \\
\bottomrule()
\caption{\small The key feedback citations collected from teachers, students,
and parents reflect common impressions across all three categories and
teachers' confidence.}
\end{longtable}

\begin{enumerate}
\def\labelenumi{\alph{enumi})}

\item
  \textbf{Program structure and way of learning.} The teachers reported
back that the program is \emph{``well-sequenced''}, \emph{``well
supported with practical demonstrations''} and \emph{``rigorous but
achievable''} and \emph{``a good blend between concepts, activities,
games and assessment''}, which helps to understand modern science in an
interesting and meaningful way.

Teachers were enthusiastic about the approach, which effectively prevent
the lessons \emph{``\ldots to get either superficial or too heavy''.}
One teacher remarked \emph{``Excellent; exposed the students to new
concepts in an active, hands-on way''.} The emphasis on outdoor time was
particularly well appreciated, with one teacher noting \emph{``The
outdoor activities were especially well received''.} On group
activities, a teacher commented that the lessons \emph{``incorporate
discussion and presentation skills through an assumption of different
roles in their groups. Students get a choice to work collaboratively
with each other or individually, or even at times to take on the role of
an observer.''}

One parent mentioned, "\emph{I am truly impressed by the wide range of
activities my child has been exposed to through this program}."
\end{enumerate}
\begin{enumerate}
\def\labelenumi{\alph{enumi})}
\setcounter{enumi}{1}
\item
  \textbf{Enthusiasm and engagement}\emph{.}Teacher responses emphasise
  a high level of student engagement and enthusiasm. Teachers reported
  that students appeared to be ``\emph{highly engaged and motivated}'',
  noting \emph{``the unprecedented growth of students numbers''.}:
\end{enumerate}

\begin{enumerate}
\def\labelenumi{\alph{enumi}.}
\item
  \emph{``Students\textquotesingle{} attendance remained high even
  though it was an after-school program ``}
\item
  \emph{"Students completed the homework which was not compulsory in the
  context''}
\item
  \emph{``Students carry over their discussions from the classroom to
  their informal conversations outside the class.}
\item
  \emph{``Students felt privileged to be included''} in the MEU program.

One teacher also noted, \emph{``We even have parents come back to
Anastasia} {[}MEU external teacher{]} \emph{with physics questions, as
children share what they have learnt with them at home''.}

Parents also indicate a high level of enthusiasm: ``\emph{At home, he
always talks about the activities he had during the lessons such as
counting rice on a chessboard and rolling balls on the elastic
space-time''} and also the impact on the attitude to science in general:
\emph{``Whenever the topic of science arises, he eagerly raises his hand
to share the ideas he has learned with MEU''}. Parents have also noticed
an increase in STEM-related discussions at home due to MEU \emph{``One
of the topics my son finds most interesting is the population of the
world. He was impressed by this unimaginable number and asked lots of
questions.''}

Most of the students expressed verbal excitement about the new way of
learning: \emph{``I like the maths of arrows. It was so cool for all of
us to push the car together! Wow! My favourite one.''}
\end{enumerate}

\begin{enumerate}
\def\labelenumi{\alph{enumi})}
\setcounter{enumi}{2}
\item
  \textbf{Effectiveness.} The teachers suggest that the program was very
  effective in achieving its learning outcomes: ``\emph{The review
  questions asked at the beginning of the sessions was a further
  indicator that students were attending and understanding what had been
  previously covered.''}
Parents described how the program benefitted their children: \emph{``My
daughter doesn't feel confident learning abstract concepts at school,
but she has very positive self-esteem with Maths for Einstein's
Universe, providing an easy and enjoyable of understanding science''.}
\emph{``Thanks to Maths for Einstein's Universe, my son
\ldots demonstrated an excellent performance}''\emph{.}
\end{enumerate}

\begin{enumerate}
\def\labelenumi{\alph{enumi})}
\setcounter{enumi}{3}
\item
  \textbf{Teachers' confidence}. The initial trials of the program
  revealed a need for additional support and assistance. A first-time
  teacher commented \emph{``the external presenter who had a greater
  depth of knowledge''} and referring to the Einstein-First person
  stated \emph{``I did need to ask her a few times whether what I was
  teaching was correct. Without her there I would have felt less
  confident.''} However, as the teachers gained experience through these
  trials, in all cases, they expressed a desire to continue implementing
  the program independently. The positive response of the teachers who
  participated in initial trials led to the creation of micro-credential
  courses (micro-credentials, 2023)as well as specialised workshops
  organized to offer ongoing support and assistance.
\end{enumerate}

\begin{enumerate}
\def\labelenumi{\alph{enumi})}
\setcounter{enumi}{3}
\item
\textbf{Future collaboration.} The schools and teachers involved in this
program were asked about future collaboration. ``\emph{I look forward to
teaching it again''; ``We are very pleased to have tried the
Einstein-First approach at our centre''}. All the trial schools indicate
that they want teach MEU on a regular basis. Students also found the
program more engaging and enjoyable than their regular classes at school
and emphasize the willingness to learn MEU. One Parent reported:
``\emph{My daughter declared that Maths for Einstein's Universe is her
favourited subject. She wants to learn science and maths in this way''};
One student said:\emph{``I want to play with toys to learn maths and
science. Spinning tops are my favourite. At school, the teacher talks
and talks about science, and then we fill in boring worksheets''}

Based on the feedback, \emph{Maths for Einstein's Universe} appears to
be a successful and engaging program for students promoting interest in
science and maths. The positive feedback on every level suggests that
the "Maths for Einstein\textquotesingle s Universe" program was a
positive and valuable educational experience for students with
remarkable outcomes. A school principal who observed the program
carefully stated \emph{``I would have not believed that it could be
taught to the target age group (7- 14-year-old students) if I had not
witnessed the astonishing comprehension that even middle primary
students have demonstrated''.}
\end{enumerate}

\textbf{Conclusion}

We have presented a hypothesis that the modernisation of mathematics
content to make it more relevant to student interests and to science,
combined with less focus on arithmetic is likely to help reduce maths
anxiety. We have suggested that maths anxiety could be reduced by
emulating the approach used in the Einstein-First program which has
shown the significant benefit of an activity-based modern curriculum in
improving teenage girls' attitudes to science (Kaur, 2020)

We have presented five modules for the \emph{Maths for
Einstein\textquotesingle s Universe} program. They focus on key
mathematical areas that a) enhance understanding of the physical
universe, and b) change the emphasise of mathematics from exactness to a
broader conceptual foundation. It is suggested that by enhancing the
understanding of modern physics concepts and associated mathematics from
the early primary school level, we will be providing students with
intellectual tools that are of value throughout our modern world while
also preparing them for exciting future learning.

We have presented evidence that MEU does indeed provide students with an
intuitive grasp of the mathematical foundations necessary for a deeper
comprehension of physical reality. Evaluations indicate that it strongly
improves student attitudes to mathematics. It is too early to determine
how it could affect future learning and maths anxiety.

We have outlined the learning progression for these modules, from Powers
to the Universe to curved space geometry along with trial results,
learning outcomes, attitudes, in-class observations, and feedback
analysis.

The initial success of the roughly 200 hours of delivered programs
indicates that Maths for Einstein's Universe effectively achieves its
teaching objectives, fosters interest and engagement in mathematics and
related science and among young students. Further quantitative results
will be presented through case studies of individual modules in
forthcoming papers.

In general, based on these preliminary observations, it is fruitful to
start learning modern physics concepts in middle primary school through
engagement with simple toys, capitalising on young learners' willingness
to play, while slowly building up their formal scientific language and
advanced abstract concepts. In order to present stronger and
statistically significant conclusions, it is necessary to conduct
observations with a larger sample size of students.

Test results for individual modules using quantitative analysis of pre
and post-tests will be published in forthcoming papers.

\textbf{Funding}

This research was supported by the Australian Research Council Linkage
Grant LP 180100859 led by Professor David Blair at the University of
Western Australia. The third author David Treagust, from Curtin
University, is a member of the Einstein-First project.~

\textbf{Acknowledgements}

The authors would like to acknowledge the contribution of all the
Einstein-First collaboration members: Marjan Zadnik, David Wood, Kyla Adams, Jessse Santos, Tejinder Kaur,Ju Li, Elaine Horne. 
We would like to thank Peter Rossdeutscher
and Howard Golden, who have enabled us to raise donation funds to
supplement our ARC Linkage funding (LP180100859)\textbf{~}that allowed
us to develop our on-line training programs. We also thank the ARC
Centre of Excellence for Gravitational-Wave Discovery (OzGrav) for their
continual support, especially in enabling us the develop school kits for
our activities. We also thank our Einsteinian Physics Education Research
(EPER) collaborators for their enthusiastic support. We are very
grateful to the West Australian Department of Education for their
support, the Independent Schools Association of Western Australia who
have facilitated many of our trials, and the Science Teachers
Association of Western Australia who have provided essential and
continuous support. We are grateful to our participating schools
principals, teachers, and students for allowing us to run the program
and for granting us permission to use their photographs and data for
research purposes. 

\textbf{Declaration of interest statement}

The authors have no relevant financial or non-financial interests to
disclose in relation to the publication of this paper. The authors also
have no conflicts of interest to declare that are relevant to the
content of this article.~
\textbf{Ethics declarations}
The participants involved in this study gave their informed consent for
this publication. The research was carried out under the University of
Western Australia Ethics approval number 2019/RA/4/20/5875.~

\textbf{References}

Bruner, J. (1976). The process of education. Harvard University.

Pitts, M., Venville, G., Blair, D., \& Zadnik, M. (2014). An exploratory
study to investigate the impact of an enrichment program on aspects of
Einsteinian physics on year 6 students. Research in Science Education,
44, 363-388.

Kaur, T., Blair, D., Moschilla, J., Stannard, W., \& Zadnik, M. (2017).
Teaching Einsteinian Physics at Schools: Part 1, Models and Analogies
for Relativity. Physical Education, 52, 065012.

Kaur, T., Blair, D., Moschilla, J., Stannard, W., \& Zadnik, M. (2017).
Teaching Einsteinian Physics at Schools: Part 2, Models and Analogies
for Quantum Physics. Physical Education, 52, 065013.

Kaur, T., Blair, D., Moschilla, J., Stannard, W., \& Zadnik, M. (2017).
Teaching Einsteinian Physics at Schools: Part 3, Review of Research
Outcomes. Physical Education, 52, 065014.

Choudhary, R. K., Foppoli, A., Kaur, T., Blair, D. G., Zadnik, M., \&
Meagher, R. (2018). Can a short intervention focused on gravitational
waves and quantum physics improve students\textquotesingle{}
understanding and attitude? Phys. Educ., 53, 065020.

Choudhary, R., Kraus, U., Kersting, M., Blair, D., Zahn, C., \& Zadnik,
M. (2019). Einsteinian Physics in the Classroom: Integrating Physical
and Digital Learning Resources in the Context of an International
Research Collaboration. The Physics Educator, 1(4), 1950016.

Foppoli, A., Choudhary, R., Blair, D., Kaur, T., Moschilla, J., \&
Zadnik, M. (2019). Public and teacher response to Einsteinian physics in
schools. Phys. Educ., 54, 015001.

Choudhary, R., \& Blair, D. (2021). All possible paths: bringing quantum
electrodynamics to classrooms. European Journal of Physics, 42, 035408.

Commodari, E., \& La Rosa, V. L. (2016). General academic anxiety and
math anxiety in primary school. The impact of math anxiety on
calculation skills. Psychology, 7(02), 196.

Luttenberger, S., Wimmer, S., \& Paechter, M. (2018). Spotlight on math
anxiety. Psychology Research and Behavior Management, 11, 311-322.

Stodolsky, S. S. (1985). Telling math: Origins of math aversion and
anxiety. Educational Psychologist, 20(2), 75-86.

Winarso, W., \& Haqq, A. A. (2019). Psychological Disposition of
Student; Mathematics Anxiety Versus Happiness Learning on the Level
Education. Journal of Physics: Conference Series, 1313(1), 012034.

Passolunghi, M. C., De Vita, C., \& Pellizzoni, S. (2017). Math anxiety
and math achievement: The effects of emotional and math strategy
training. Frontiers in Psychology, 8, 1629.

Organisation for Economic Co-operation and Development. (2013). PISA
2012 Results: Ready to Learn (Volume III): Students' Engagement, Drive
and Self-Beliefs. Paris: OECD Publishing.
\url{http://dx.doi.org/10.1787/9789264201170-en}.

Lee, J. (2009). Universals and specifics of math self-concept, math
self-efficacy, and math anxiety across 41 PISA 2003 participating
countries. Learning and Individual Differences, 19(3), 355-365.

Ashcraft, M. H., \& Krause, J. A. (2007). Working memory, math
performance, and math anxiety. Psychonomic Bulletin \& Review, 14(2),
243-248

Thuneberg, H. (2016). Hands-On Math and Art Exhibition Promoting Science
Attitudes and Educational Plans. Journal of STEM Arts, Crafts, and
Constructions, 1(1), 1-6.

Valentini, M. (2019). The math game: How motor activity and the use of
own body can help in mathematical learning. Systematic review.
International Journal of Environmental and Science Education, 14(11),
769-786.

Isbister, K. (2018). Scoop! A Movement-based Math Game Designed to
Reduce Math Anxiety. In Proceedings of the 2018 Annual Symposium on
Computer-Human Interaction in Play (pp. 347-357).

Maass, K., Geiger, V., Romero Ariza, M., \& Goos, M. (2019). The Role of
Mathematics in Interdisciplinary STEM Education. Journal of Science
Education and Technology, 28(6), 685-697.

Blaizer, J. (2020). Strategies for reducing math anxiety. Journal of
Mathematical Behavior, 57, 100754.

Kaur, T., Blair, D., Choudhary, R. K., Dua, Y. S., Foppoli, A.,
Treagust, D., \& Zadnik, M. (2020). Gender response to Einsteinian
physics interventions in school. Physics Education, 55(3)

Ashcraft, M. H., \& Krause, J. A. (2007). Working memory, math
performance, and math anxiety. Psychonomic Bulletin \& Review, 14(2),
243-248

Mahajan, S. (2018). The exponential benefits of logarithmic thinking.
American Journal of Physics, 86, 859.

Johnson-Laird, P. N. (1994). Mental models and probabilistic thinking.
Cognition, p. 189-209.

Demi. (1997). One Grain of Rice. New York, NY: Scholastic.

Williams, R. E., Baum, S., Bergeron, L. E., Bernstein, N., Blacker, B.
S., Boyle, B. J., Brown, T. M., Carollo, C. M., Casertano, S., \&
Covarrubias, R. (2000). The Hubble Deep Field South: Formulation of the
Observing Campaign. The Astronomical Journal, 120(6)

Rueckner, W., \& Titcomb, P. (1996). A lecture demonstration of single
photon interference. American Journal of Physics, 64(8), 986-991

Blair, D. (2023). Playing Einstein. Big Picture Factory. ISBN
987-1-922781086

Blair, D., Burman, R., \& Davies, P. (2022). Uncovering
Einstein\textquotesingle s New Universe: From Wallal to Gravitational
Wave Astronomy. World Scientific Publishing Company.

Blair, D., \& Cody, G. (2012). The History of Time: Universe. Gravity
Discovery Centre.

Popkova, A., Adams, K., Boublil, S., Choudhary, R. K., Horne, E., Ju,
L., Kaur, T., McGoran, D., Wood, D., Zadnik, M., Blair, D. G., \&
Treagust, D. F. (2021). Einstein-First: Bringing children our best
understanding of reality. World Scientific Publishing 58(3), 345-370

Blair, D. G., Kaur J., Popkova A. (2022). School activities to celebrate
the centenary of the Wallal Eclipse Expedition. Teaching Science, 68(3),
23-27.

https://www.einsteinianphysics.com/micro-credentials/

\end{document}